**The pale blue dot: using the Planetary Spectrum Generator to simulate signals from hyper realistic exo-Earths**

Kofman[1,2], Villanueva[1], Fauchez[1,2], Mandell[1], Johnson[3], Payne[1], Latouf[1,4], Kelkar[1,5]
1: NASA Goddard Space Flight Center, 8800 Greenbelt Rd, Greenbelt, MD, 20771, USA
2: Physics Department, College of Arts and Science, American University, 4400 Massachusetts Ave NW, Washington, DC 20016
3: Dept. Physics and Astronomy, University of Las Vegas, 4505 S. Maryland Pkwy. Las Vegas, NV 89154, USA
4: George Mason University, 4400 University Drive, Fairfax, VA, 22030, USA
5: IISER, Indian Institute of Science Education and Research, Dr. Homi Bhabha Road, Pune, 411008, India

**Abstract**
The atmospheres and surfaces of planets show tremendous amount of spatial variation, which has a direct effect on the spectrum of the object, even if this may not be spatially resolved. Here, we apply hyper-realistic radiative simulations of Earth as an exoplanet comprising thousands of simulations and study the unresolved spectrum. The GlobES module on the Planetary Spectrum Generator was used, and we parameterized the atmosphere as described in the modern-earth retrospective analysis for research and applications (MERRA-2) database. The simulations were made into high spatial resolution images and compared to space-based observations from the DSCOVR/EPIC (L1) and Himawari-8 (geostationary) satellites, confirming spatial variations and the spectral intensities of the simulations. The DISCOVR/EPIC camera only functions in narrow wavelength bands, but strong agreement is demonstrated. It is shown that aerosols and small particles play an important role in defining Earth's reflectance spectra, contributing significantly to its characteristic blue color.
Subsequently, a comprehensive noise model is employed to constrain the exposure time required to detect $O_2$, $O_3$ and $H_2O$ as a function of varying ground and cloud cover for several concept observatories, including the habitable worlds observatory. Cloud coverage enhances the detectability of planets in reflected light, with important consequences for the design of the future HWO. The HWO concept would require between 3-10 times longer to observe the studied features than LUVOIR A but performs better than the HabEx without a starshade. The codes, routines, and the noise models are made publicly available.

# 1. Introduction

Future observatories that aim to characterize Earth-like exoplanets around sun-like stars are now being conceptualized, as the LUVOIR, HabEx and the Habitable Worlds Observatory (Decadal Survey on Astronomy and Astrophysics 2020 (Astro2020) et al. 2021; Gaudi et al. 2020; The LUVOIR Team 2019). Part of these efforts consist of simulating the reflection spectra of planets, considering ground coverage, atmospheric variability, and aerosol coverage. Realistically, to simulate these correctly, it requires taking into account different angles of incidence, refraction, and numerous spectroscopic considerations. Furthermore, the performance of the telescope or telescope concepts will have to be assessed. Leveraging data from the actual Earth and simulating this as an exoplanet provides a test bed to assess the detectability of ground and atmospheric phenomena. Examples of these include varying surface coverage as a function of the phase (e.g., vegetation red edge (Arnold et al. 2002; Fujii et al. 2013, 2011; Kawahara & Fujii 2011; Livengood et al. 2011; Tinetti et al. 2006; Wang & He 2021; Woolf et al. 2002) and ocean glint (Livengood et al. 2011; Lustig-Yaeger et al. 2018; Robinson et al. 2014, 2010; Ryan & Robinson 2022; Trees & Stam 2022, 2019; Vaughan et al. 2023)).

Several efforts have reproduced full-disk images of Earth, either verified using Earth's light reflected by the moon (Earthshine, (Pallé 2018; Pallé et al. 2003; Qiu et al. 2003; Turnbull et al. 2006) or by spacecraft observations (Bessho et al. 2016; Burt & Smith 2012; Livengood et al. 2011). Observations of the entire disk provide an opportunity to understand the effects of several physical phenomena, which can be modeled, but these need to be validated using real observations. These include modeling of the terminator, the effect of various clouds and other aerosols on the brightness and the detectability of surface coverages, the potential visibility of vegetation, and ocean glint. The opportunity to do this using real spacecraft and instrumentation data provides a level of realism that is essential to validate numerical simulations.

In an extended mission of a cometary probe, Deep Impact, the instrumentation was turned to the Earth-Moon system and imaged in the UV-vis, as well as globally integrated in the near-infrared (Livengood et al. 2011). Earth was observed during three intervals of approximately 24h from March 18 to 19, May 28 to 29, and June 4 to 5 in 2008. A number of studies investigated the images and light curves from the EPOXI mission (Cowan et al. 2011, 2009; De Cock et al. 2022; Fujii et al. 2013, 2011; Merrelli et al. 2019; Robinson et al. 2011). Accurately reproducing the light curves requires consideration of three-dimensional projections of the reflected solar light on the Earth's surface and the propagation of radiation through the atmosphere at each location. In (Robinson et al. 2011), the authors reproduced several full Earth disk images taken by the instruments on board the Deep Impact flyby spacecraft in the extended EPOXI mission. The HEALPix, or *Hierarchical Equal Area isoLatitude Pixelization division,* scheme was used to combine pre-calculated reflectance spectra, interpolated over solar/observer incident/reflection angles, to produce the surface coverage base map of the Earth. Atmospheric properties were averaged from several sources and evaluated using the SMART (Spectral Mapping Atmospheric Radiative Transfer) radiative transfer code (Crisp 1997; Meadows & Crisp 1996). Four different types of clouds were parameterized: ice clouds at 0.33 bar (close to 8.5 km) and liquid water clouds at 0.85 bar (close to 1.5 km), both either described as totally opaque or moderately opaque. The cloud abundances and types were adopted from MODIS and superimposed on the combined surface-atmosphere spectra. The Terran World Spectral Simulator (TSS, Merrelli et al., 2019) follows a similar approach where the planet's sphere is divided in a number of tiles which are subsequently filled with a linear combination of surface and atmospheric templates. Each combination of atmospheric and surface template is evaluated separately using LBLDIS RTM routines, which is a combination of the Line-by-Line radiative Transfer Model (LBLRTM) and Discrete Ordinates Radiative Transfer (DISORT) models (Clough et al. 2005; Stamnes et al. 1988; Turner 2005), see section 3.2 of (Merrelli et al. 2019) for more details.

More recently, images from the Deep Space Climate Observatory (DSCOVR), equipped with the Earth Polychromatic Imaging Camera (EPIC) have been used to study spectra of the full Earth disk. DSCOVR observes the fully illuminated Earth disk from Lagrange point 1, taking high-resolution images approximately 10-20 times per day, at 10 narrow spectral channels throughout the UV-vis (317.4 to 780 nm). The DSCOVR/EPIC images and spectra provide an excellent opportunity to study full-disk images of Earth.  Time series can be used investigate what surface properties would be retrievable from integrated disk spectra, similarly to the EPOXI observations, but over much longer time-series (Aizawa et al. 2020; Fan et al. 2019; Gu et al. 2021; Jiang et al. 2018; Kawahara 2020). Analysis of the time series have demonstrated that it would be possible to disentangle changing atmospheric and surface properties using principal component analysis or more sophisticated techniques. Finally, in Gu et al., 2022, it was demonstrated the spectral response of the different spectral channels could be reproduced (i.e. average errors lower than 20%) by their Earth Spectrum Simulator (ESS). The simulations leveraged input from the DSCOVR/EPIC composite repository a (NASA/LARC/SD/ASDC 2017), which derives the cloud coverage density from the same images as Gu et al., reproduced.

The effects of clouds on the detectability of biomarkers in the visible/NIR using LUVOIR was investigated by (Kawashima & Rugheimer 2019; Wang et al. 2018). Both studies demonstrated the detectability of $O_2$ was enhanced over low cloud coverage as the albedo is significantly higher. Wang et al. investigated the detectability $CH_4$, $H_2O$, $O_2$ and $CO_2$ separately, considering one molecule at a time at different heights of cloud coverage (high – corresponding to 12 km, low at 4 km versus no clouds), finding that a low cloud cover left the strongest atmospheric signals over a highly reflective surface and thus provided best chances for detectability. Kawashima and Rugheimer came to similar conclusions using a 1-dimensional radiative transfer model. In this study, clouds were represented by continuum absorbing/emitting layers, and the cloud coverage varied by taking the weighted average of cloudy and clear spectra of Earth observed at a phase angle of 90 degrees. In the infrared cloud coverage decreases flux emitted from the Earth, thus reducing the detectability of molecular signatures (e.g. see Figures and 3 and 4 in (Kitzmann et al. 2011)).

Our study adds to the previous works by performing full 3-dimensional radiative transfer calculations of Earth atmospheric models, relying on 3D atmospheric models for clouds (both ice and liquid water), $H_2O$ and $O_3$ profiles. The model consists of 144 longitudinal by 91 latitudinal by 72 vertical bins. Ground coverage is compiled from monthly average satellite-based observations. Full radiative transfer calculations, at the appropriate incidence/emission angles, are performed for each of the profiles on the visible side of the planet. Clouds and other aerosols are considered using the PSGDORT multiple scattering scheme, based on optical properties derived from Mie calculations. We describe recent advances on the GCM module of the Planetary Spectrum Generator, GlobES, that enable the 3D simulations (Fauchez et al., 2024, in review). GlobES can ingest GCM files either converted from netCDF files by the supplied python codes on the GitHub page (https://github.com/nasapsg/globes), or from custom made input climate files, as is done in this work. For this study, high spatial resolution Earth data is used to simulate how future telescopes would observe Earth as an exoplanet, and what aspects of its surface (*i.e.*, snow, ocean, glint) and atmosphere (clouds) would be observable. Furthermore, a full telescope/instrument noise model is applied to simulate required exposure time with several concept observatories like the Habitable Worlds Observatory.

## 2. Methods

Simulating realistic observations of exoplanets requires consideration of several physical, astronomical, and technological aspects. Many teams are working on constraining what spectroscopic signature molecules of life, or biomarkers, could be targeted to constrain whether an exoplanet hosts life. Similarly, climate states are simulated relevant to specific planetary states, such as tidally locked planets, or investigating the photochemistry under different types of stellar hosts. The subsequent step in investigating what the diverse range of exoplanets may look like is figuring out how much of these signals would be visible by future and current telescopes. In order to simulate this, accurate models of the astrophysical scene, the telescope, and the instrument/detector have to be utilized.

For the simulations in this study the Planetary Spectrum Generator is used. In particular, the local version, hosted on a Docker virtual machine is employed. The local version releases the online limitation of the number of simulations GlobES can employ from a maximum of 200 to the ~6500 simulations used here. The different aspects of the calculations will be explained in this section, but for more details see the web resources (https://psg.gsfc.nasa.gov), the PSG handbook (Villanueva et al. 2022), or the original manuscripts describing the tool (Villanueva et al. 2018). For specific details on the calculations, the configuration files, which are a collection of the commands sent to PSG to perform the simulations, are made available. All of the parameters are listed and described in the online repository (https://psg.gsfc.nasa.gov/helpapi.php#parameters). The most relevant settings for these calculations will be explained in the sections below.

At the center of the calculations is the radiative-transfer (RT) code PUMAS, which is the RT model in the Planetary Spectrum Generator. It uses correlated-k tables to model the absorption of gases, based on the latest versions of several popular molecular and atomic databases, such as HITRAN and HITEMP (Gordon et al. 2022; Rothman et al. 2010). For the absorption of water not directly attributed to line-by-line transitions, and thus not captured in these databases, but instead described in the MT_CKD model, the CIA parameterization is applied as described in (Kofman & Villanueva 2021). Finally, additional UV absorption from the Mainz database (Keller-Rudek et al. 2013) and (Serdyuchenko et al. 2014) are considered as well for this investigation.

The surface of the planet is modeled employing several scattering methods including a Lambert model, which considers that light is homogenously scattered across all directions, with the amount of light being reflected back being directly proportional to the illuminated area. The single scattering albedo of the different surfaces considered here are adopted from the United States Geological Survey, which are available in PSG (Kokaly et al. 2017). For the simulations in this study, surface properties are mixed using areal linear mixing (*i.e.*, the proportional sum of the constituents). For oceans, in addition to the Lambertian diffuse reflection, specular reflection from solar light is included. In the case of reflections on a body of water, the roughness of the water enlarges the area at which sunlight is directly reflected much beyond the size of the solar disk on the Earth. This effect is typically described as glint and increases with the speed of the wind in the area. Cox and Munk (1954) have done a detailed study analyzing the glint strength as a function of the wind speed, and their parameterization is adopted here. Additional refinements, such as the shadow effect of facets at high incidence angles, are taken from (Jackson & Alpers 2010). The full description of the implemented Cox-Munk model is given in the PSG handbook (Villanueva et al. 2022). For the glint calculations, a speed of 8 m/s was adopted here. This was found to best reproduce the spot size and intensity of the glint feature in the different space-based observations. Lower wind speeds results in a bright but smaller reflectance spot. This simplification deserves further study, and wind speed vectors are available in MERRA-2 could be potentially ingested in future investigations.

To consider the effects of aerosols and Rayleigh scattering in the atmosphere, the radiative transfer calculations need to take into account multiple scattering processes. A brief, conceptual description of the problem will be given here, for more details see the PSG resources, or (Stamnes, 1986; Stamnes et al., 1988 or the DISORT documentation). In addition to radiation being traced from the source, through the atmosphere, and to the detector, there are several ways in which light may scatter that need to be accounted for. Not only is plane parallel light considered, but diffuse light from sources *within* the layers is considered. This requires the quantification of the amount of light that is scattered at several different angles from within the layer. Of particular relevance here is that the intensity of the light that is scattered varies strongly with the angle, a value that is captured by the asymmetry parameter *g*. What is ultimately required for the multiple scattering calculations is a method that keeps track of the absorbed, emitted, and scattered light, budgets all of these phenomena, and yields a balanced solution. The multiple-scattering solver in PSG is called PSGDORT, which is based on the widely used and validated DISORT package. PSGDORT solves the RT task by reducing the multiple scattering to a series of numerical approximations in the form of differential equations, which are ultimately solved by a series of matrix operations. As one can imagine, the approximations can be progressively more complex by increasing the number and size of the equations describing the scattering functions. The number of equations, or 'streams' is encoded by Nmax, the 'size' of the equations is indicated by the number of Legendre polynomials 'Lmax'. Guidance on the choice of these parameters is given in the PSG handbook (Villanueva et al. 2022). Briefly, to include Rayleigh scattering, Nmax=1 and Lmax=2 are sufficient. This corresponds to 2 (2x Nmax) additional streams, their equations described by 2 Legendre terms (Lmax). For the simulations here, Nmax=4 and Lmax=60 are chosen, which we found to be sufficiently accurate but not overly demanding. For more details see the PSG handbook (Villanueva et al. 2022). The calculation of the wavelength dependent extinction coefficients and scattering albedo are calculated using Mie routines that are made publicly available on our GitHub page.

### Ingesting Global Climate Model netCDF files into GlobES

Two files are used when calculating spatially resolved calculations from PSG. The configuration file is the central paradigm which is used in all calculations in PSG. A comprehensive list of all the settings/commands of the config file is found at https://psg.gsfc.nasa.gov/helpapi.php, here we will briefly go over the commands used for these calculations. For GlobES calculations three-dimensional global climate models are ingested in a binary format, the details of the construction will be described below. Alternatively, the binary information can be directly added to the configuration file as well.

Within the configuration file, the following settings are relevant for the study here.

- Several data sources are combined to make the 3-dimensional array containing 5 different fields for ground coverage (144 longitudinal by 91 latitudinal bins, or at a 2.5-by-2-degree resolution). The atmosphere consists of 144 by 91 by 72 bins. The vertical layering that is adopted provides a temperature and pressure at the center of each bin, so the vertical spatial resolution is dynamic.

- The surface properties adopted here are constructed using ground coverage from the MODIS database. The *moderate resolution imaging spectroradiometer* consists of two observatories, and the team releases high-resolution maps of the ground coverage of Earth every year (MCD12C1). For simplicity, the ground coverage was divided in 5 categories: ocean, snow, soil, forest, and grass, covering several of the sub-types in the database. We bin the resolution from a 0.05 by 0.05 degree to 2.0 by 2.5 (from 7200 by 3600 to 144 by 91) and save the fraction of the ground coverage in a GlobES climate file. For snow coverage, monthly average data is obtained from another MODIS source (MOD10CM). These records supply snow coverage for the entire Earth, aside from the polar night-side. For our purposes, these areas are assumed to be snow covered. Data for October 2022 was missing so we adopted the same information from October 2021 instead. Sea-ice coverage is taken from the National Snow & Ice Data Center (Meijer et al., 2021). These are reported using an equal-area scalable Earth, or 'EASE-Grid' (see https://nsidc.org/data/user-resources/help-center/guide-ease-grids, or Brodzik et al., 2012) projection of the Northern and Southern hemisphere (*i.e.*, the poles are at the center of a round map). The EASE-Grid projections need to be converted to the regular latitude-longitude projection map that can be ingested in PSG. Elements of the NSIDC '*NSIDC Polar Stereographic Projection lon/lat conversion: polar_convert*' package were adopted to recast the grids. An additional longitudinal correction was needed to align Antarctic to the correct position when the code '*polar_convert*' was used. As the north pole area is at the edge of the instrument's coverage, it is masked in the ice-coverage data of this area. For these purposes, the area was assumed to be covered with (permanent) sea ice. The Antarctic land coverage mapping is slightly different between the lon/lat MODIS land-coverage and the EASE-Grid south pole coverage. In cases where there was an inconsistency between the two, for consistency the EASE-Grid's coverage was adopted, since its projection provides more accurate coverage in that region.

- The cloud record is taken from the Modern-Era Retrospective analysis for Research and Applications (MERRA-2) database. MERRA-2 includes several dozens of atmospheric parameters and contains record of these since 1979. ). Specifically, PSG works with the M2I3NVASM component, which provides assimilated meteorological fields (pressure, temperature, water vapor, ozone, and water ice clouds) from the surface to ~80 km (72 layers) with a cadence of 180 minutes, and spatial resolution of ~0.5 degrees (576 x 361). The MERRA2 data are binned down to the same resolution as the ground-coverage information. As well as the abundance of cloud particles, the cloud coverage fraction is reported in MERRA-2, which is used to construct the simulations as described below. The water and the water ice clouds are modeled assuming a particle size distribution centered around 5 and 100 μm respectively. The aerosols are defined in the relevant fields in table 1. With the 'scl' tag in the abundance units, PSG is signaled that it will have to look for a profile, either in the layer-by-layer atmosphere description, or the GCM if the name of the aerosol is given in the ATMOSPHERE-GCM-PARAMETERS field. The ATMOSPHERE-GCM-PARAMETERS field deserves some extra consideration as this string details the entire format of the GlobES climate file that is loaded into PSG, so it is crucial that this is filled correctly. Its contents are covered in detail in Fauchez et al 2024.

- The atmosphere is simulated using 3-dimensional parameters for $H_2O$ and $O_3$ from MERRA2, and for $N_2$, $O_2$, $H_2O$, $CO_2$, $O_3$, and $N_2O$ we considered profiles as reported for the US standard atmosphere. Surface reflections, Rayleigh scattering, refraction, collision induced absorption are included in the simulations. The images are extracted from the simulations by adopting the intensities in the spectral bins closest to the RGB channels for the respective images. Spectra are simulated at a resolving power of 70, which is one of the possible configurations of the future Habitable Worlds Observatory.

Table 1: Overview of the parameters used in the PSG configurations file that are relevant for the calculations done here. The full configuration file will be made available in the Supplementary Information (SI_1_cfg_GCM_EPIC.txt). For more information, the PSG website contains the full list of fields and descriptions (https://psg.gsfc.nasa.gov/helpapi.php#parameters).

| Configuration field name | Field value |
|---|---|
| **SURFACE-NSURF** | 5 |
| **SURFACE-SURF** | Ocean, Snow, Grass, Soil, Forest |
| **SURFACE-TYPE** | Albedo_GSFC, Albedo_GSFC, Albedo_GSFC, Albedo_GSFC, Albedo_GSFC |
| **SURFACE-MODEL** | Cox-Munk, 8.0 |
| **SURFACE-PHASEMODEL** | ISO |
| **ATMOSPHERE-AUNIT** | scl, scl |
| **ATMOSPHERE-AEROS** | SeaSalt, Water, WaterIce |
| **ATMOSPHERE-ATYPE** | AFCRL_Seasalt_HRI, AFCRL_Water_HRI, Warren_ice_HRI |
| **ATMOSPHERE-ASIZE** | 0.078, 5.0, 100.0 |
| **ATMOSPHERE-ASUNI** | um, um, um |
| **ATMOSPHERE-CONTINUUM** | Rayleigh, Refraction, CIA_all, UV_all |
| **ATMOSPHERE-GAS** | N2, O2, H2O, CO2, O3, N2O |
| **ATMOSPHERE-TYPE** | HIT[22], HIT[7], HIT[1], HIT[2], HIT[3], HIT[4] |
| **ATMOSPHERE-ABUN** | 78.1, 20.9, 1 400, 1, 0.3, |
| **ATMOSPHERE-UNIT** | pct, pct, scl, ppm, scl, ppm |
| **ATMOSPHERE-GCM-PARAMETERS** | 144, 91, 72, -180, -90, 2.5, 2.0, Surf_Ocean, Surf_Snow, Surf_Soil, Surf_Forest, Surf_Grass, Temperature, Pressure, Water, WaterIce |

## Inclusion of clouds and aerosols

The inclusion of clouds is more challenging than the molecules and processes described above for two reasons. First, clouds have strong gradients in opacity with altitude and have a non-linearity in the brightness of the reflected light and the total abundance of the particles. The second reason is that the spatial descriptions of the clouds are incomplete as only the total water mass (liquid or ice) and the cloud fraction in the bins are given. In order to capture the spatial inhomogeneities of the cloud distributions, we separated the simulations into two cases. The first set of simulations are cloud free (Fig. 1, first panel), while In the second simulation, each bin is calculated at the maximum cloud opacity by scaling mass fraction by the inverse of the fraction. For example, if the bin has a 20% cloud coverage and an abundance of 2e-3 g/kg, we assume these 10e-3 g/kg clouds are present throughout the bin. This results in a nearly fully clouded planet (second panel). In a post-processing step, the resulting images of both simulations are combined using a 2-dimensional projection the cloud abundance (third panel). The 2D projection is obtained by combining the vertical profiles of the cloud fraction and abundance using the following equation:

$$cc_{[0-1]} = \frac{\sum CC_p * water_p}{\sum water_p},$$

where $CC$ indicates the cloud coverage factor of the now 2D map, and $water_p$ is the abundance of water in kg/kg. Finally, the two cloud coverage maps for ice and liquid clouds are added together, cloud fractions are capped at 1, and the 2D map is projected onto a 3D globe matching the orientation of the simulations using the PyGlobES `map2d_to_globe` function. Although the majority of the cloud features are reproduced, some of the more tenuous clouds are under described, or sometimes over reproduced. This deserves further investigation but is probably primarily related to the limitations of the climatological database in capturing the historical state of the atmosphere, and the simplifications in the particle size distribution of the clouds assumed here.

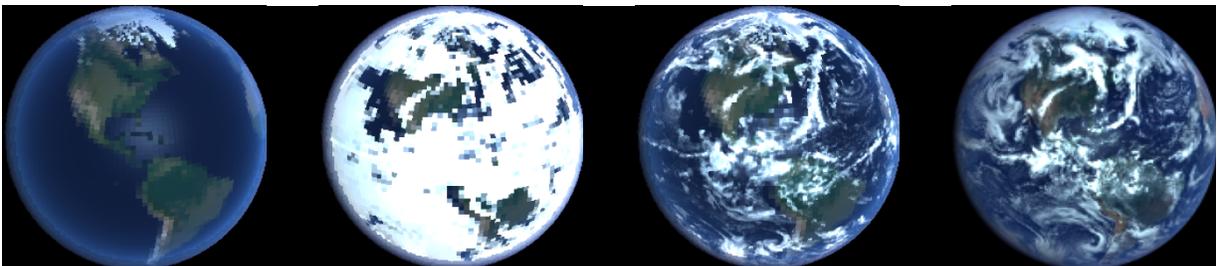

*Figure 1: Progression of the simulations, combining the cloud-free simulations (left most panel), with the fully cloudy simulations (panel two), to yield the simulations using the cloud fraction map (panel three). The last panel shows the DSCOVR/EPIC observation that is being reproduced. Observations and synthetic images show very similar dynamic range, absolute flux calibration and coloring. No special or arbitrary scaling or calibrations were applied to the model, nor the data.*

## Inclusion of additional scattering due to aerosols

Earth's atmosphere contains a considerable amount of non-water aerosols, which have a significant effect on the radiative transfer and the energy budget of the atmosphere (Charlson et al. 1992; Hansen et al. 2023; Kremser et al. 2016; Li et al. 2022). The scattering and absorption of these aerosols adds a large magnitude of uncertainty in projections of global warming, possibly offsetting between one third or half of the expected global warming (Hansen et al. 2023). The sources are diverse and range from direct emission of sea salt from the oceans, dust from lose soils, particles volcanos, to biomass burning (natural or anthropogenic), and (photo)chemically produced from sulfates, nitrates, and organic components. Unlike the greenhouse gases, the precise role of aerosols is hard to quantify because they have limited lifetimes in the atmosphere, may provide local heating or cooling at different heights in the atmosphere, and interact strongly as cloud condensation nuclei, which can both seed clouds on the one hand, but deplete water on the other hand. The primary effect of the aerosols can be ascribed to a general Mie scattering behavior, resulting in a re-distribution of Solar light (Bohren & Huffman 1983; Massie & Hervig 2013; Villanueva et al. 2022). Many of the particle are significantly less than a micron in size resulting in a general increase in the scattering of shorter wavelength light. (Malinina et al. 2018) for instance investigated the sizes of aerosol in the stratosphere, using retrieval methods on Earth limb observations to constrain the particle size and distribution, yielding ranges around 0.1 µm between 18 and 32 kilometers.

In this work, we approximated the effects of these various aerosols, excluding water ice and water clouds, with a single Mie scatterer. A small particulate aerosol was added to the simulations of the atmospheres corresponding to an optical depth around 0.15, which is the average observed on Earth (*e.g.*, see Figure 4 in Zhang and Reid, 2006). Since optical depth is typically measured at a particular wavelength in the optical (in the case of Zhang and Reid at 550 nm), and the scattering effect aerosols is strongly wavelength dependent, there remains a measure of uncertainty in this approximation. If we assume small particles (*i.e.,* particle size distribution centered at 0.08 µm) as reported in the literature described above, an optical depth of 0.15 at 550 nm is relatively consistent with an isoabundance profile of 5 ppb kg/kg. Furthermore, and considering that sea-salt is a dominating aerosol in many regions of our planet, we assumed this composition for the particles. In Figures 2 and 3 the importance of the addition of aerosols is explored both from the perspective of the color perception of the Earth as well as the spectral radiance of the Earth observed at a distance.

## 3. Results

The day side images of Earth are compared to DSCOVR/EPIC observations (Figure 2, top panels). DSCOVR is in a Lissajous orbit around L1, where it takes dayside images of the Earth approximately every hour. The PSG/GlobES simulations are shown in the top row, the EPIC observations are shown in the second row. MERRA-2 reports the state of the atmosphere from its global circulation re-analysis model every three hours, the observations are picked as close to these moments as possible to minimize differences, and the appropriate solar and observer latitude/longitude combination is picked from the DSCOVR observations. The DSCOVR/EPIC images are binned down from the 2048x2048 pixels by a factor 10 to aid the comparison to the models. From the images it is clear that the ground cover simulations accurately capture the different features of the Earth. The Saharan desert is simplified, but that can be attributed to the fact that a single ground cover is chosen to represent soil. Most of the main cloud features are seen in the simulations, although the density of some of the patchiness is not fully captured by the model. This is partly due to the resolution of the simulations, where the finer details of the coverage cannot be represented by the 2-by-2.5 degrees pixels. One may also wonder whether the MERRA-2 data fully captures cloud coverage in all of the areas of the globe accurately, since is it is a re-analysis is based predominantly on weather data from the more populated areas. Alternatively, the adaptation from MERRA-2s clouds fraction for radiation and mass fraction of cloud ice/liquid water may require some refinement. This certainly warrants further investigation.

The second panel of Figure 2 shows the simulation of images obtained from the Japanese Himawari-8 weather satellite. This observatory is in a geostationary orbit and takes images of the Earth every 10 minutes and thus provides an excellent test bed to compare images of Earth at crescent phases. The Himawari observations clearly show the sun reflecting on the ocean, a phenomena that has been proposed to enable the remote observation of water bodies on exoplanets (Lustig-Yaeger et al. 2018; Robinson et al. 2010; Vaughan et al. 2023). This effect will be investigated in more detail below when the full spectra of the simulations are discussed. Comparing the two sets of images (simulations in the top and the Himawari-8 observations in the bottom panel), we see that the model, similarly to the EPIC observations, reproduces most of the cloud and surface features relatively well. The deserts of Australia are again not well represented, but particularly for the Himawari-8 images it proved challenging to find images where the gamma and relative strength of the colors appeared closer to true color (*i.e.*, it appears that blue and red for instance are enhanced compared to the colors seen in the EPIC observations). Human eye color perception is anything but straightforward, and with the generation of RGB images information about the intensity of the colors is lost. For our images, the wavelength dependent responses are calculated in spectral radiance units (W/sr/m$^2$/μm), which takes into account the Solar emission spectrum. Recent work has revisited earlier images of Neptune and Uranus and discusses in detail the 'translation' of linear CCD values to 'average' color perception, see Irwin et al., 2024, but also see Miller et al., 2016. In the case of the Himawari-8 images, the color scaling (*e.g.*, the intensity distribution of red pixels) are very different from the EPIC images at the same time and approximate location, complicating the comparison, because the Himawari-8 images are processed using non-linear scaling relationships for each channel. The focus of the comparison is instead on the spatial variations. The glint spot appears roughly the same size, which is remarkable as a single wind speed is used for the entire planet in the simulations. The wind speed is an essential parameter in the simulation of ocean glint in the methods used for these simulations (Cox & Munk 1954; Villanueva et al. 2022). For more discussion see the glint section below.

Finally, the simulations of the pale blue dot, generated from the full 3-dimensional simulations are shown in the bottom panel of Figure 2, the atmospheric conditions corresponding to the time of observation (February 14, 1990, 04:48 UTC), as well as the original image as imaged by Voyager. Note that although the Earth covered approximately 1/6$^{th}$ of a pixel on the Voyager camera from the distance, the light is nonetheless spread of several dozens of pixels because of the instrument point spread function.

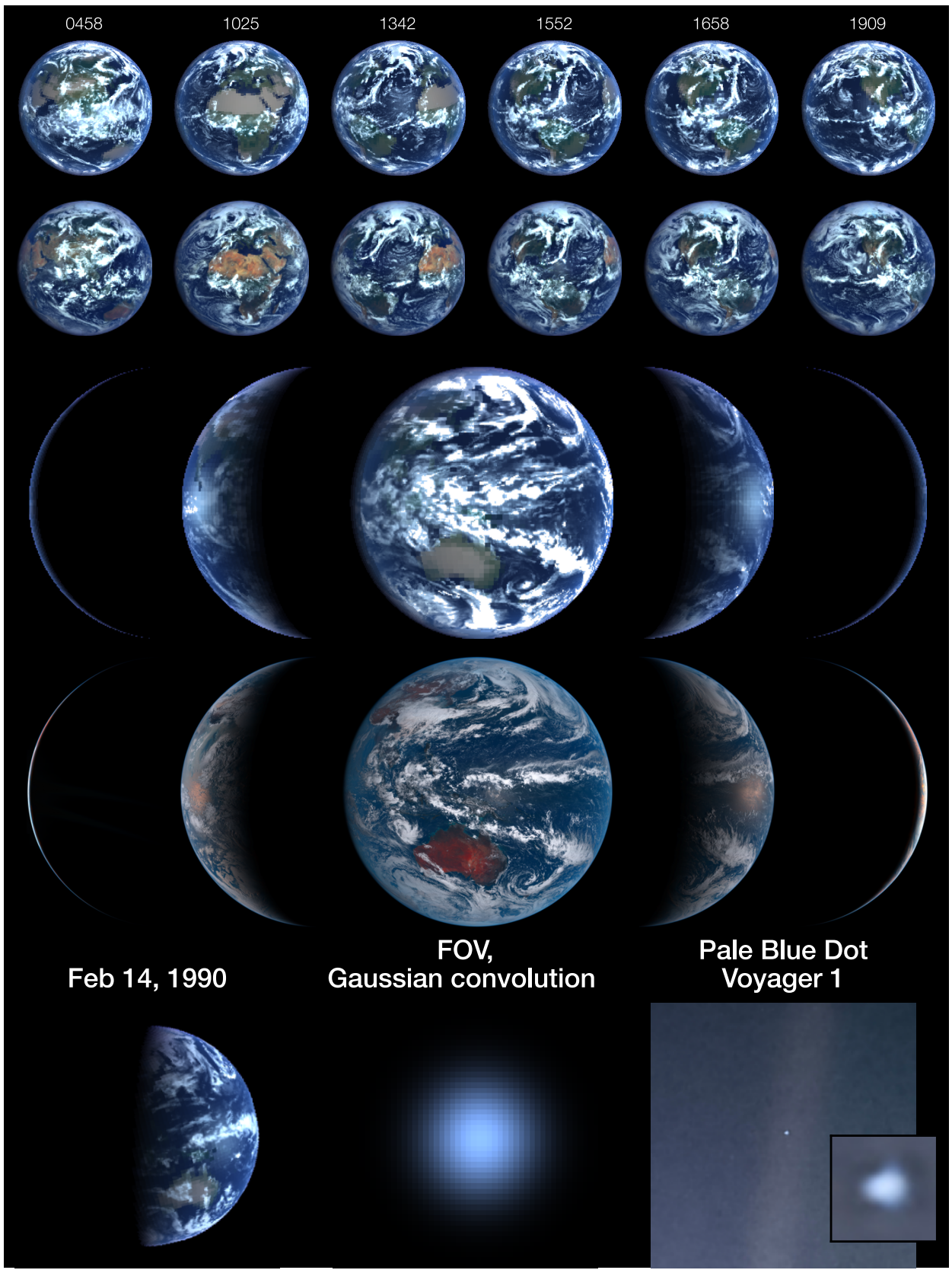

*Figure 2: **Top panel**: Simulations (top row) and DSCOVR/EPIC observations of Earth (bottom row) at spring equinox in 2022. UTC times of the simulations and observations are indicated in above the figures. **Second panel**: Simulations (top row) and Himawari-8 observations of the Earth from geostationary orbit (longitude 140.7). **Third panel**: in order of appearance: 3D simulation of the Earth at the time and orientation of the Voyager 1 spacecraft. At the distance of 40 AU, the pixel size of 5.3e-4 degree corresponds to approximately 6 times the Earth's diameter. This is unresolved by Voyagers 1's cameras and results in the blurry image on the right. The discussion section elaborates more on the generation of color from spectral radiance.*

## Calibration of the DSCOVR/EPIC spectra and spectral comparison to our models

Even though the EPIC instrument contains relatively narrow spectral channels, see Table 2 for the full-width half-maximum, it provides the best resource available to us to verify the spectral intensities of the simulations done here. In (Herman et al. 2018; Marshak et al. 2018), the calibrations of the spectral bands of EPIC are described. The channels are calibrated such that the corrected CCD counts correspond to reflectance values of the nadir instruments MODIS *Aqua* and *Terra* satellites in the near visible and near infrared (Marshak et al. 2018), and to the Suomi National Polar-Orbiting Partnership Ozone Mapping and Profiler Suite (Suomi NPP – OMPS) in the UV (Herman et al. 2018; Li et al. 2017). Briefly, after dark-current subtraction, flat fielding, and corrections for stray light, the CCD counts can be scaled to match the reflectance values of the OMPS and the MODIS observatories. Finally, since our values are reported in spectral radiance (W/μm/m$^2$/sr), the reflectance values are multiplied by the top-of-the-atmosphere spectral radiance values to match the simulations. Using these calibrations, the EPIC observations are used both to verify the disk-integrated fluxes, as well as constrain how well local variations are captured in the PSG simulations. The top of the atmosphere solar fluxes in PSG are based on the Kurucz 2005 G-type stellar template for the 0.4-1 um range and the LISIRD database (LISIRD 2005) for the <0.4 um points. Note that the brightness of the EPIC spectral bands and the PSG simulations agree well across all wavelengths, with the exception of the simulations 0142, 0458, and 0709. The first simulation in the top panel of Figure 2 shows a bright patch of clouds that is not seen as prominently in the EPIC images, which would explain this discrepancy. This patch is also visible in the 0142 and 0709 simulations (not shown).

Within the literature, different definitions of albedo and its variation are used and at this point it is important to elaborate on the definition used here. Geometric albedo is the radiance of a planet as seen at phase zero relative to the flux intercepted by its cross-section. A flat white disk would have a geometric albedo of 1.0, while a white Lambertian sphere a geometric albedo of 2/3, and in the case of a Lommel-Seeliger surface (e.g., the Moon) a value of 1/8. For other phases, we adopt the term albedo to refer to the radiance relative to the same cross-section as for phase zero, similarly to the definition of albedo in Cahoy et al. 2010. Please keep in consideration, that this definition of *apparent* albedo differs from that in Qiu et al. 2003, in which albedo is defined relative to a Lambert sphere at that specific phase, while our definition is always relative to a flat white disk at phase 0

*Table 2: Spectral bands of the DSCOVR/EPIC instrument and their spectral calibration based on the solar incidence radiation calculated using PSG. The two spectral bands at 688 and 784 nm respectively correspond to specific absorption bands of O2 which are not resolved at the resolving powers studied here and are excluded in the comparison.*

| Wavelength [nm] | 317.4 | 325 | 340 | 388 | 443 | 551 | 680 | 780 |
|---|---|---|---|---|---|---|---|---|
| FWHM [nm] | 1.0 | 1.0 | 2.7 | 2.6 | 2.6 | 3.0 | 1.6 | 1.8 |
| $K_\lambda$ [-] | 1.216e-4 | 1.111e-4 | 1.975e-5 | 2.685e-5 | 8.340e-6 | 6.660e-6 | 9.300e-6 | 1.435e-5 |
| Spectral radiance [W/μm/m²/sr] | 221.57 | 251.49 | 304.87 | 329.14 | 603.75 | 600.22 | 477.24 | 381.96 |

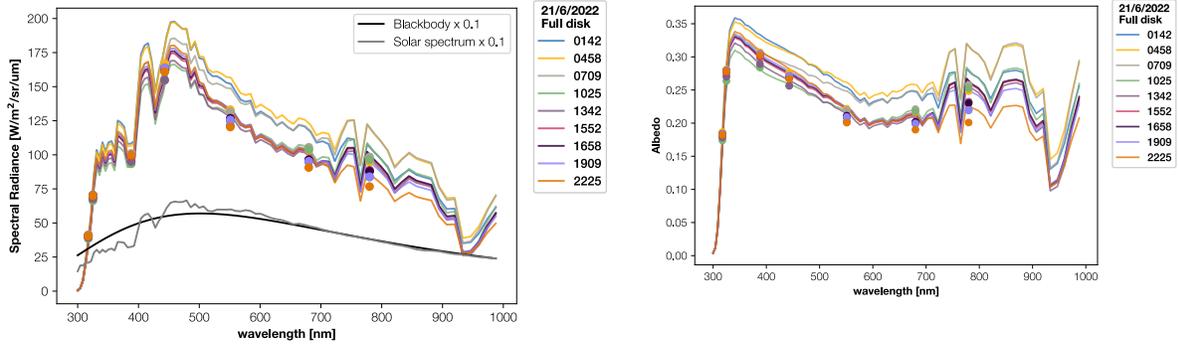

*Figure 3: Comparison of the spectra obtained from this study with the calibrated DSCOVR/EPIC bands in units of spectral radiance (left) and albedo (right). The values for albedo of the EPIC observations are obtained directly from their calibrations, whereas spectral radiance values are obtained by multiplying these values with the spectral radiance of the sun at the top of the atmosphere. In the figure on the left, the solar spectrum and the 5777K black body spectrum are shown, scaled by a factor 0.1.*

## Application to Exoplanet observations

In PSG, there are three different modes of calculation that can be explicitly selected for exoplanet observations. Figure 4 shows the cases and their corresponding phase, defined here for clarity. So far, we have described the planet as if the full disk is visible, what is described in PSG as the eclipse mode. For planets where the orbital inclination is close to 90 degrees this will result in a secondary eclipse (the star is assumed not to be in the field of view for the calculations above). Maximum separation on the sky is seen at quadrature (P). The last case considers the planet being directly in front of the star or transiting the host star. These modes can be selected under the 'change object' button on the home screen by the S, P, and T buttons respectively, which will modify the ephemeris accordingly. It should be noted that the simulations are all fully constrained by sub-solar and sub-observer latitude and longitude however, and that these modes really only are there to help the user select the appropriate geometries.

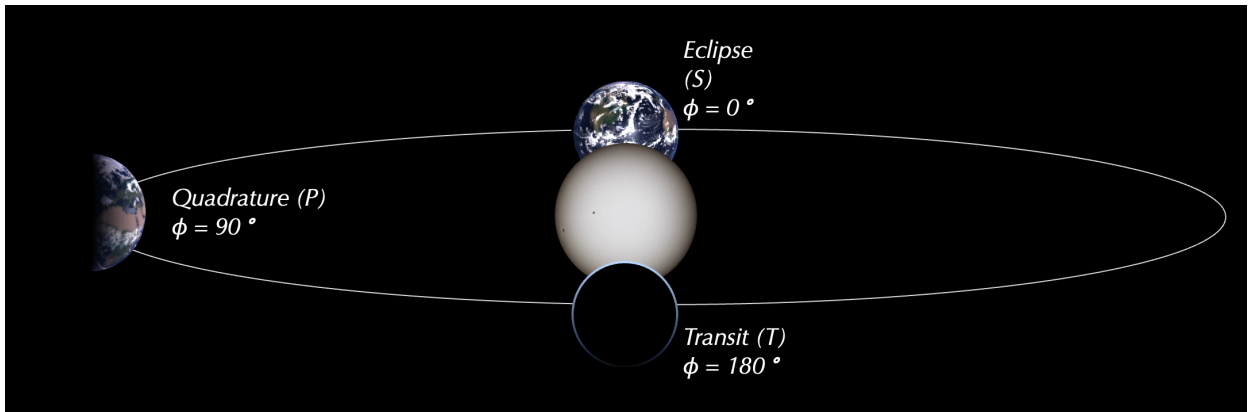

*Figure 4. Definitions of the different orbital phases used in this work, as adopted from the descriptions Planetary Spectrum Generator. This is by no means the only work that describes these phases as such, but in the interest of clarity it is included here.*

The previous section showed a number of examples of variation in spectroscopic signals from disk-integrated spectra of the Earth demonstrated full disk, or the eclipse phase. In the case of exoplanet observatories of Earth like exoplanets with a telescope like the concepts HabEx, LUVOIR or HWO, the coronagraph instrument requires a minimum angular separation on the sky (inner working angle) that essentially prohibits full disk observations in all but the closest systems. Eclipse observations are too close to the star, so the planet will be observed in quadrature or close to this where the planet-star angular separation is maximum. For the rest of the simulations in the work, the simulations are performed with the planet at quadrature.

The following sections focus on the detectability of spectroscopic features in the spectra from the simulated observations. Nine different times during summer solstice of 2022 are simulated. These track the difference in the orientation of the planet, and thus the effect of varying ground and cloud coverage. Several spectroscopic features from ground coverage are considered to be of interest as signals indicative of vegetation (the vegetation red edge) or the specular reflection of sun light, indicative of large water bodies could be targets for the study of the habitability of an exoplanet (Lustig-Yaeger et al. 2018; Robinson et al. 2014; Ryan & Robinson 2022; Schwieterman et al. 2018). The detectability of both of these effects, both with and without clouds are highlighted in this section as well.

## Variation in spectral features as a result of planetary rotation

### Diurnal variation 2022 – 06 – 21

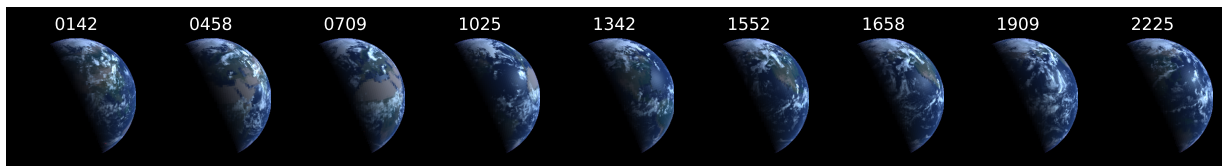

**Cloud-free planet**          **Cloudy planet**

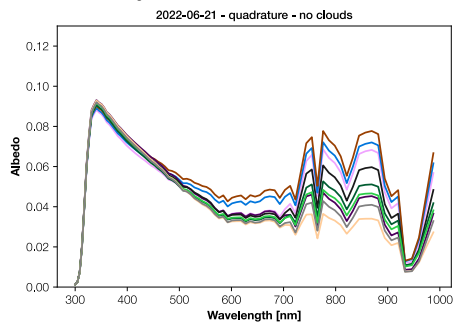          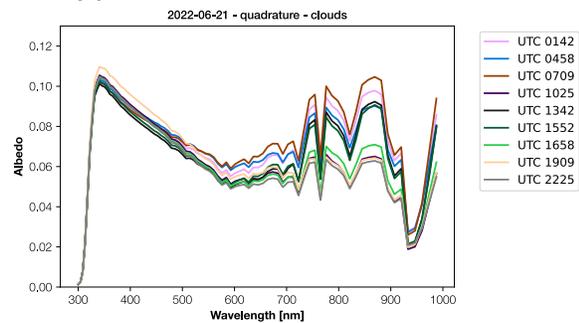

**Detectability, no clouds**          **Detectability, with clouds**

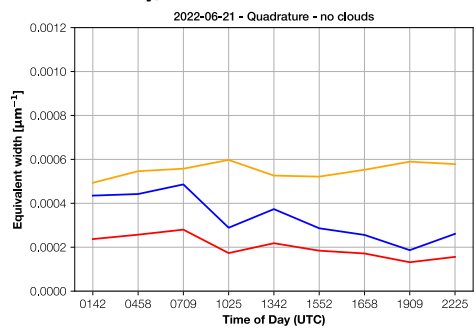          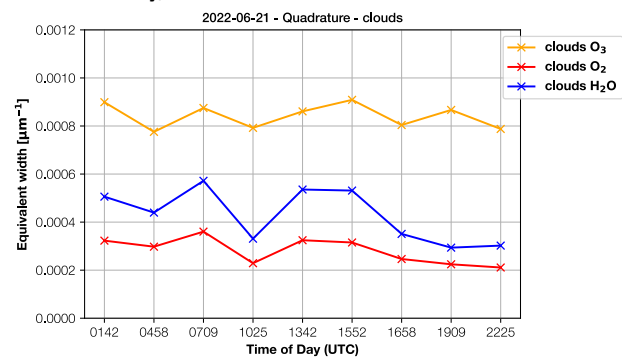

*Figure 5 Simulations of the albedo of Earth during the summer solstice in 2022, calculated using ~6500 sub-calculations at a resolving power of 70. On the left, spectra of the calculations without clouds are shown, the combined cloudless and cloudy*

*simulations are shown on the right (see text for more information). The images in the top are generated from the spectra in units of spectral radiance, which is subsequently converted to albedo for the middle panels. The intensity of the molecular bands is integrated into equivalent width, which is shown as a function of UTC in the bottom plots.*

Figure 5 shows the simulations of Earth at quadrature in its 2022 summer solstice at several different times. As described in the previous section, the images and the simulations consist of two parts, with and without clouds. Note that the difference in the albedo is roughly a factor three between the full disk simulations and the simulations at quadrature. This highlights the importance of 3D effects when considering the simulations of planets at various orientations. The difference cannot simply be scaled by the Lambertian integral, since clouds, Rayleigh and aerosol scattering are strongly non-isotropic. As the cloudless atmospheres are available, they are separately analyzed to constrain the effect of the Earth's surface and clouds in the atmospheres separately. In the top panel the orientations are shown, with the corresponding times in the titles. The middle row shows the disk averaged spectra of a cloudless atmosphere on the left, and with clouds included on the right. If no clouds are considered, the difference in the spectra between the various orientations is significant, and the vegetation red edge is clearly visible. When clouds are included however, this effect is not as pronounced since clouds hide the effect as there are additional photons reflected by the clouds. Similar to what is observed in the DSCOVR observations (e.g. (Gu et al. 2021), the change in the cloud deck dominates the spectral variation. Whereas in the cloud free situation, the variations in the maximum at 340 nm are small, in the cloudy case the variations are much larger. Interestingly, this 340 nm maximum is the strongest at UTC 1909, where the Pacific Ocean is mostly visible. At UTC 2225 the Pacific Ocean dominates the view as well, but cloud cover is significantly less. Also, note that maximum of the albedo in the cloud-free cases is at 340 nm, whereas in the cloudy cases more light is seen at 860 nm in some of them. The variations in cloud and ground coverage reveal a complex interplay between the surface and cloud heterogeneity and how much this affects the albedo varying throughout the spectral range.

The bottom row of Figure 3 shows the integrated spectral bands of $O_3$, $O_2$ and $H_2O$ in the spectra shown in the top panel. These values are obtained by selecting the range at which the molecules are known to absorb light and removing the spectral feature by a linear interpolation between the start and the end of the feature. $O_3$ features are integrated between 514 and 673 nm, $O_2$ between 750 and 790 nm and the $H_2O$ feature between 787 and 857 nm are integrated. There is a stronger $H_2O$ feature between 900-1000 nm, but less photons are available in this range, and this is close to the edge where current detectors lose sensitivity. Note that these boundaries are chosen to include the spectral features in the simulations done here, and not necessarily correspond to the best integration ranges at other resolving powers. The linear interpolation between these points allows one to estimate the signal strengths that originate from these molecules, and to study their variation between the different orientations and at varying cloud coverage. Although this is as a relatively simplified approach to quantifying the signal from the molecules, it identifies the maximum amount of signal that could be recovered from an observation and thus serves as a first estimate to constraining the detectability. Alternatively, one could calculate the spectra with and without the molecules of interest and calculate the detectability from the difference. For two reasons this was not done here. First of all, it would at least double, or possibly quadruple, the number of simulations that would have had to be performed. The second motivation is that it is unlikely that we would be able to obtain the full spectroscopic signal originating from the molecules in real observations. For the narrower signals from $H_2O$ and $O_2$, constrains are expected to be more accurate. $O_3$ however, as it is a broad signal, might be much more degenerate with other contributions in this range. For these reasons, the detection strengths should be considered as upper limits.

It is observed that the detectability of the different species in the cloudless case varies strongly depending on which part of the Earth is visible, but that this variation is somewhat muted when clouds are included. Interestingly, clouds enhance the detectability of $O_2$ and $O_3$ quite strongly but have a more convoluted effect on the visibility of $H_2O$. The effects of the clouds on the detectability of $H_2O$ mostly relates to the height of the different types of clouds, and how they relate to the abundance profile of $H_2O$ in the atmosphere. We calculated the average abundance of $H_2O$, $O_3$ and the liquid and ice clouds from one of the GCM snapshots. For $H_2O$, 90% is found below ~ 4 km, so depending on the height of the clouds most of the molecules may actually be below the cloud deck, which reach the 90% abundances at 5 and 8 km from the surface respectively. $O_2$ and $O_3$ are much more abundant higher in the atmosphere and thus the detectability is mostly enhanced (90% cumulative abundances at 14 and 32 km respectively). Considering that these are the Earth's average values for one particular time, and there will be a strong variation of these values with temperature, not to mention variations in the abundances of $H_2O$ and $O_3$ throughout the atmosphere, this gives an indication of the complex interplay between clouds, their altitude, and the detectability of the spectral features.

## Oceanic glint and its detectability

Specular reflection from liquid water has been studied as a potentially detectable signature of liquid water on Earth-like exoplanets (Lustig-Yaeger et al. 2018; McCullough 2006; Robinson et al. 2010). Specular reflection originates from light reflecting off liquid water bodies and is often called ocean glint. Wind interactions with the surface break up the flat water surface, and light will reflect of the wave crests further out from the direct reflection (*i.e.,* the point where the sun-surface angle is equal to the surface observer-angle). Cox and Munk provided a numerical method to simulate glint considering a gaussian distribution of waves based on parameterization of the wind speed (Cox & Munk 1954). Although including glint increases the total amount of flux returned from the planet, the spectral dependence of this effect is limited increasing slightly for longer wavelengths (Ryan & Robinson 2022). The spectral dependence originates in the long atmospheric path-lengths the specularly reflect light experience through the atmosphere. As the effect is very broad, and changes very gradually with wavelength its effects appear degenerate with an increase in cloud coverage, variations in the aerosol abundance, ground coverage, or planet size if that is not constrained. For the wavelength ranges covered in this investigation, confidently identifying ocean glint in the crescent phase of an exoplanet containing an ocean would appear very challenging. Temporal variations, *i.e.* the 'blinking effect' such as described in (Lustig-Yaeger et al. 2018) may break some of the degeneracies, but not the one with clouds. The temporal variations also may not be accessible due to low photon fluxes.

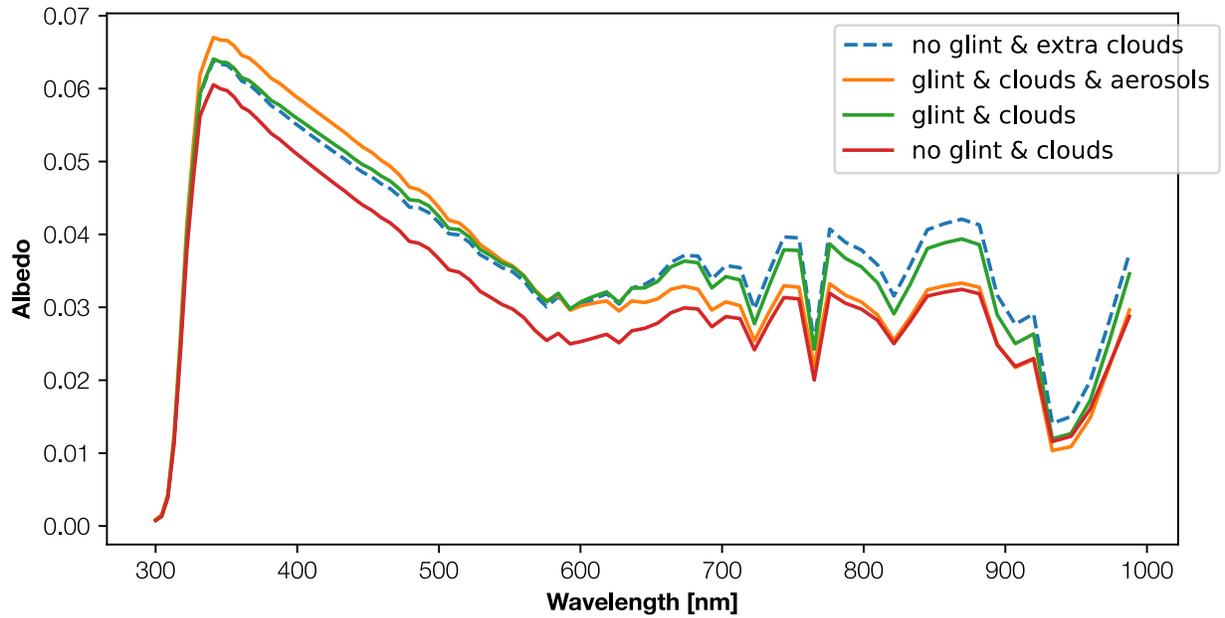

*Figure 6. Simulations with different phenomena enabled to study the relative contribution of ocean glint to the total albedo. For these simulations, the Himawari-8 observations were adapted (4th image in the middle panel of Figure 1).*

## 4. Noise simulations

Constraining how the observations of the planet and its atmosphere as simulated above would be detected in a telescope requires considering careful consideration of each of the elements within the light path of the observatory, aside from the astrophysical noise sources (shot noise, zodiacal light). This includes the wavelength dependent throughput of the mirror-instrument-detector path, efficiency of the detector, as well as noise sources from the observatory (aberrations, speckle noise, detector sources). In PSG, modeling of the performance and noise levels of a telescope is modeled using a set of eleven parameters, of which some are wavelength dependent when a coronagraph instrument is being modeled. For more details see (Checlair et al. 2021; Kopparapu et al. 2021; Saxena et al. 2021). The setting of each of the observations used in this study are shown below. An example configuration file for the Habitable Worlds Observatory is made available the supplemental information, which follows the requirements posted by the decadal survey. Where information was missing, the same values for LUVOIR-B were adopted, as HWO was recommended to be an off-axis segmented telescope. A python adaptation of the PSG noise simulation package is made available as well (https://github.com/VincentKofman/PySGnoise). This code enables noise simulations after simulations are already performed, saving considerable time when many simulations are performed in GlobES. The code should also be instructive for the way PSG calculates the different noise fluxes, and provides an adaptable python code for general purpose sensitivity calculations. For the most up-to-date noise simulations the user is directed to PSG however.

For the simulations here, the coronagraph is modeled by multiplying the flux from the star by the contrast ($10^{-10}$), and subsequently the rest of the telescope parameters (wavelength dependent throughput, detector characteristics, etc.). The signal from the planet is multiplied by the wavelength-and-inner-working-angle dependent coronagraph throughput before the other telescope parameters are considered. For these simulations, no specific bandpass is considered, as ideally these would be designed or placed optimally to constrain the spectral feature considered. For the case of $O_3$, a 20% bandpass might not be sufficient to constrain the broader feature that is targeted. Although this could be easily adopted in the noise simulation package that is used, a critical assessment of the bandpass placements and widths is part of other efforts (BARBIE, see Latouf et al., 2023a, 2023b). In the BARBIE studies, Bayesian retrievals are applied, which allows for an investigation between the possible degeneracies between spectroscopic features.

Table 3. The relevant telescope and instrument settings for the different observatories studied here are listed below. The HWO is currently only a recommendation from the 2020 astrophysics decadal study, so the settings adopted are the requirements of the observatory from this study.

| Description | PSG-config parameter | HabEx (SSI) | HabEx (HCG) | LUVOIR A | LUVOIR B | HWO req.* |
|---|---|---|---|---|---|---|
| Diameter [m] | GENERATOR-DIAMTELE | 4 | 4 | 15 | 8 | 6 |
| Contrast | GENERATOR-TELESCOPE1 | 1e-10 | 2.5e-10 | 1e-10 | 1e-10 | 1e-10 |
| Inner working angle vs. throughput | GENERATOR-TELESCOPE3 | 2.49 L/D | 2.49 L/D | 9.32 L/D | 4.97 L/D | 3.73 L/D |
| Temperature optics T & I [K] | GENERATOR-NOISEOTEMP | 270 | 270 | 270 | 270 | 273 |
| Throughput T & I (incl. QE) [AVG] | GENERATOR-NOISEOEFF | 0.7 | 0.21 | 0.27 | 0.28 | Same as LUVOIR B |
| Emissivity T & I | GENERATOR-NOISEOEMIS | 0.1 | 0.1 | 0.1 | 0.1 | 0.1 |
| Read noise (incl. CIC) [e-] | GENERATOR-NOISE1 | 0.008 | 0.008 | 0.008 | 0.008 | 0 |
| Dark rate (incl. CRs) [e- /s] | GENERATOR-NOISE2 | 3e-5 | 3e-5 | 3e-5 | 3e-5 | 3e-5 |
| Lower limit [um] | GENERATOR-RANGE1 | 0.450 | 0.702 | 0.702 | 0.702 | 0.702 |
| Upper limit [um] | GENERATOR-RANGE2 | 0.975 | 0.858 | 0.858 | 0.858 | 0.858 |
| Nr pixels | GENERATOR-NOISEPIXELS | 10 | 10 | 10 | 10 | 10 |

The setting GENERATOR-TELESCOPE3 requires some additional explanation. It provides the coronagraph throughput as a function of the angular separation. The values quoted here are used by PSG to calculate GENERATOR-NOISEOEFF. Other relevant parameters GENERATOR-TELESCOPE2 indicating the exo-zodi level, the number of pixels on the detector (GENERATOR-NOISEPIXELS), NOISETIME, and NOISEFRAMES, have been left out of the table as these will be discussed later. Light from extrasolar zodiacal dust adds an additional source of photon flux, and it is highly dependent on the targeted system (for instance see (Kammerer et al. 2022)). For the studies here, zodiacal dust is not considered since the focus is on the variations of the planet and its cloud cover, and on the difference between the telescopes. Zodiacal dust would similarly affect all simulations. For the different telescopes, a larger mirror would help mitigate the effects as the spatial resolving power is higher. The online PSG tool does include a full dust model, both for the Solar System dust as exo-zodiacal dust.

To obtain the signal-to-noise ratios for each of the telescope concepts, the simulations of the observations are performed using a python-based adaptation of the noise model that is in PSG. This python module was written to make specifically for the applications here, and to study the effects of the different telescope parameters. It also allows simulations of the noise based on the albedo of the planet and as such it decouples the noise simulations from the spectral calculations. This provides several advantages: It significantly speeds up noise calculations for calculations that rely on sub-calculations (like the ~6500 used here), it does not require additional simulations with PSG, and it allows for a more flexible exploration of the detectability of spectral features as a function of the telescope or geometry of the system. The dominant source of noise for these observations is photon noise or also called shot noise, originating from the stochastic nature of photons arriving to the detector. Other sources of noise come from the detector (the dark count rate), telescope emission (negligible in this wavelength range), and background fluxes from zodiacal and exozodiacal dust (the latter are ignored for this study). Each of the points within the range can be combined, using the equation below. This equation considers the signal originating from the signature of interest ($m_{s,k}$), the spectrum without the signal ($m_{w,k}$), and the noise ($\sigma_k$) all at the relevant wavelengths. The strength of the signal is then obtained by taking the square root of the sum of the squares:

$$S/N_{s-w} = \sqrt{\sum_k \left(\frac{m_{s,k} - m_{w,k}}{\sigma_k}\right)^2}$$

The noise package considers as an input the telescope (with the corresponding specifics), the spectral feature under investigation, the planet (its reflectance spectrum, geometrical parameters), and the required target S/N for the detection of the feature. In the next section, the required observational time for these noise levels using the different telescope concepts is constrained. The results are shown in Figure 7, which consists of 3 panels, showing the times required to detect $O_3$, $O_2$, and $H_2O$, at a S/N ≥ 5 for the five different telescope concepts, both with and without clouds. Each of the nine different UTC times shown in Figure 5 are analyzed.

### $O_3$ – 514 to 673 nm

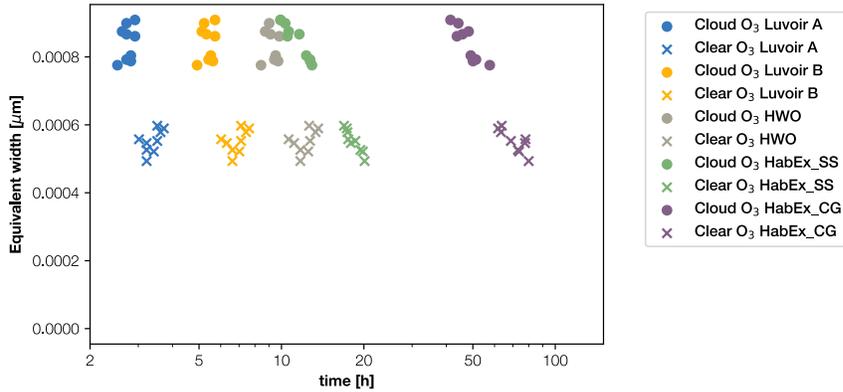

### $O_2$ – 750 to 790 nm

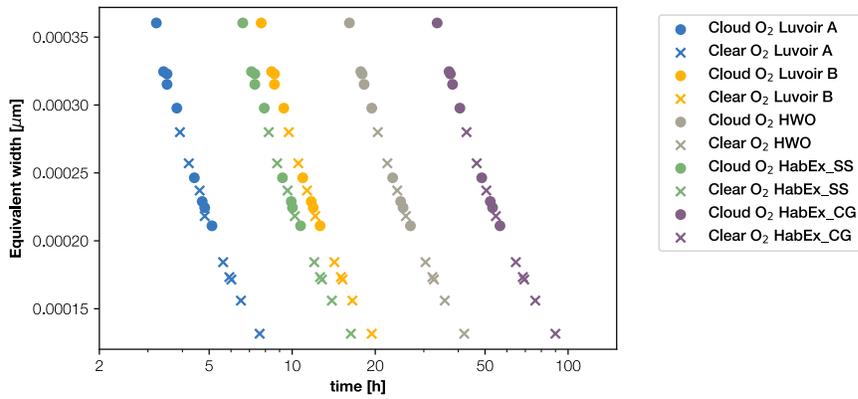

### $H_2O$ – 787 to 857 nm

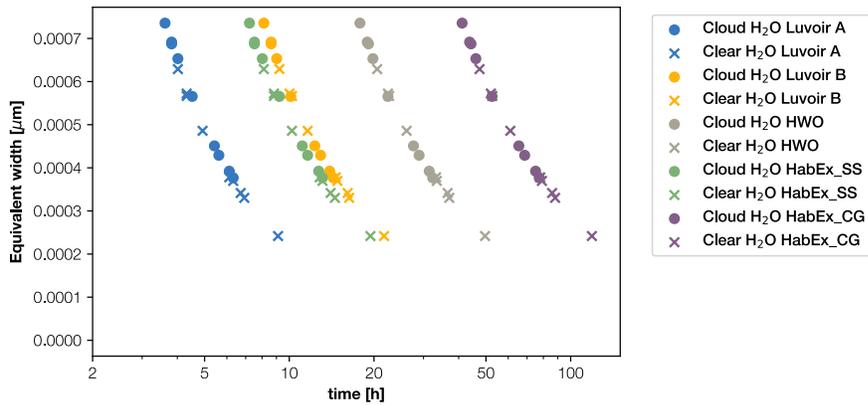

*Figure 7. Exposure time requirements to reach a S/N of 5 for the integrated band for Earth seen from 10 parsec for the five different telescope concepts studied here. The full sets of parameters describing the concepts is given in Table 3. The y-axis shows the equivalent width at the times indicated in the figure. Both the cloudless and simulations with cloud coverage are shown.*

For most of the cases the observational times are too long to consider the roughly three-hour cadence realistically. The observations span over the time of several orientations and average out the spectra. To highlight the effect of the variation in the albedo studied here though, the results are presented case-by-case. It should also be noted that with the relatively unconstrained observatory parameters the exposure time simulations should be interpreted more in a relative than an absolute sense.

Comparing the different telescope concepts, it is evident that the size of the photon collecting power (*i.e.*, primary mirror size combined with throughput) is the strongest factor that drives the detectability for spectroscopic features of an Earth-like planet at 10 parsecs. Except for the $O_3$ case, the detectability of the spectroscopic features follows a relatively simple square-root dependence of time, or rounded curve in log-linear space. Particularly for the case of $O_3$ though, there are counter intuitive dependencies, where in some cases a stronger signal needs more integration time to reach the S/R of 5. These appear to be effects of the depth of the feature versus the total amount of signal returned (a deeper feature at an overall lower albedo is harder to detect). Particularly for the cloud-free case this results in an irregular dependence on the feature strength, which partly originates from the coronagraph throughput function in the LUVOIR instrument (which is used in both LUVOIR telescopes, as well as the HWO). In the case of the HabEx instrument the throughput is more constant with wavelength in this range, resulting in exposure time requirements more intuitive.

Of particular interest for this study is the variation between the observational times that are required for the same observatory, but for different planetary orientations/cloud coverage fractions. For the case of the detectability of $O_2$ in the case of the LUVOIR A concept, a range of 3-10 hours can be required in order to obtain an S/N of 5. 10 hours is also sufficient to constrain the same signal of $O_2$ in a moderately bright case with the LUVOIR B concept, a much smaller telescope. When the HWO observatory is considered, the variation ranges between 15 and 45 hours. The order of the duration of the different observations also varies whether a cloud-free or a clouded atmosphere is considered. This seems highly relevant as well, as this effect decouples the detectability from ground coverage and more towards the state of the cloud deck.

## 5. Discussion

The methods of calculating cloud coverage from the abundance and the coverage fraction proved to be a satisfactory approach to simulating clouds. The simulations in Figure 2 where the EPIC images are reproduced show the main features are reproduced well. Differences are particularly at very high latitudes, where the simulations underestimate the cloud density. As the MERRA-2 data is a retrospective GCM, it deserves further study whether this is due to the estimations of the clouds in the database or due to the way clouds are treated in the simulations.

The 3D simulations highlight several aspects of radiative transfer that cannot be represented accurately by calculations where a disk-averaged profile of the planet is assumed. The effect is particularly strong for aerosols, but relevant for Rayleigh scattering as well. In particular for future studies where Earth-like planets would be observed at quadrature, this is of considerable importance since the flux may well be over- or under-estimated if non-isotropically scattering processes are incorrectly accounted for by 1D codes. Although the case in Figure 5 is of course only a single series, the magnitude of the variation is relevant, and changes up to 50% in the albedo are seen. Similarly, the addition of clouds to the 3D simulations provides large variations in the albedo of the planet. As with all simulations, the simulations should be used more as examples of a real exoplanet than ground-truth spectra. However, the effects of the different spatial variations on the detectability of the spectral features should serve as a good benchmark for what is to be expected in real three-dimensional planets. For ground coverage, the relevance of 3D simulations mostly applies to specular reflection or ocean glint. Even though its effects may be degenerate with or even masked by clouds or ground cover, it should still be considered since approximately 70% of the planet is covered in water and this effect can cause a significant brightening of the albedo of the planet under optimal conditions.

Realistically, however, using 3D simulation tools for all studies of the detectability of specific biomarkers, atmospheric effects, or changes in the ground coverage of the planet will not be possible. The challenge is now to devise a method that agrees well with the full 3D simulations but does not necessarily require thousands of radiative transfer calculations. One such method is the 'sub-disk sampling' as is described in Chapter 2 of the PSG handbook (Fauchez et al., 2024; Villanueva et al., 2022; Saxena et al. 2021). This method weights the incidence and emission angles in concentric circles and performs a number of calculations as specified by the sub-sampling number $N$. For non-symmetric effects such as cloud cover, however, an alternative approach will have to be found as these contributions currently require two independent simulations. The study and methods described in this work may provide a benchmark for 1D or disk-sub sampling routines to verify the fluxes obtained.

Studies at higher spectral resolution than the ones performed here may be of interest as well. More spectral information is available at higher resolutions, so a future observatory may be designed to be capable of resolving powers up to 200 or higher, versus the 70 used in this study. There is a strong trade-off between the higher resolving power and the increase in required observational time, as the same amount of photons get spread out over more pixels. Spectral contrast (i.e., the difference in flux between the continuum and absorption features) may be enhanced, which would particularly benefit retrievals as the signals can be more uniquely attributed to specific phenomena or molecules. For the study here though, it is the integrated band strength, or equivalent width, that is analyzed and expressed with respect to the noise. As long as the features do not overlap, the integrated band strength does not significantly change with the resolving power. This was tested for the $O_2$ feature that is investigated here. For this comparison, a simplified 1D simulation was done to get an estimate on what variation to expect. The spectral band was integrated at several resolutions. The comparison of the equivalent [$\mu m^{-1}$] width at the various resolving powers (R) is as following: R70: 3.4e-4, R100: 4.1e-4, R140: 3.7e-4, R200: 3.5e-4, R500: 3.1e-4, and R5000: 3.0e-4. The variation corresponds to about 13% in the R70 vs. R5000 case, versus 36% in the R100 vs R5000 case. The variation originates from the placements of the spectral bins and how this propagates through the radiative transfer. A more detailed investigation is beyond the scope of this study; this comparison mostly serves as an estimate for what variation one might expect for a lower-resolution simulation. To demonstrate the utility of 3D simulations and study the detectability as a function of telescope specifics, the R=70 case is sufficient.

The noise calculations proved to be insightful for the detectability of the spectral features of interest, and how these change as a function of the telescope concept studied. Since many studies have focused specifically on the HabEx and LUVOIR concepts, the comparison shown gives an indication of how well the recommended HWO would perform compared to the earlier studied concepts. Particularly as HWO is now in the pre-design phase, the information provide here - and the noise calculator where the telescope features can be modified one-by-one – may prove insightful for other studies. It should be noted that the instrumental details of the observatories are best estimates, and the exposure time calculations should be considered as guiding. One particularly interesting aspect is that it is possible to apply the noise calculation tool to earlier simulations: it calculates the planet's flux based on the albedo. As it is purely Python-based, this reduces the need to run any simulations in PSG. From the study here it is apparent that there are strong variations in the expected detection times required, depending on how the target planet is simulated.

Full 3D simulations can in the future be applied to other planets or different climate states for Earth. Considering the non-isotropic effects from aerosols and the surface, there is a significant variation to be expected from simulating cloudy or hazy planets, potentially improving the expected yield of future observatories such as the Habitable Worlds Observatory.

## 6. Conclusions

Including clouds and other relevant aerosols in the atmosphere positively affects the detectability Earth-like exoplanets, as determined from these 3D simulations. In particular, molecular features are significantly easier to detect when realistic clouds are considered. Looking specifically at the different molecules studied here, $H_2O$ behaves slightly different from $O_2$ and $O_3$. As $O_2$ and $O_3$ are abundant above the clouds, their detectability is enhanced under cloudy conditions since more photons are reflected of the cloud decks than of the typical ground coverage. The detectability of $H_2O$ is typically enhanced, but since its confined to the lower parts of the atmosphere, its spectroscopic signals may also be obscured by clouds. For future studies, it is important to consider realistic cloud coverages and heights to constrain the detectability of spectroscopic signals. No atmosphere in the Solar System is without clouds and assuming exoplanets atmospheres are cloud free is unrealistic. Furthermore, the python-based noise model easily allows the trade study between different telescopes, instruments, and planetary systems. It has been demonstrated that the performance of the HWO is on a similar level as the LUVOIR and HabEx concepts, but that optimization strategies should be considered to further define the observatory. Realistic simulations of planetary atmospheres are an important component of this.

**Acknowledgements:**
VK would like to acknowledge the help of Tanvi Deshmukh in the partial testing of the PyGlobes python module. This work was supported by the GSFC Sellers Exoplanet Environments Collaboration (SEEC) and the Exoplanets Spectroscopy Technologies (ExoSpec), which are part of the NASA Astrophysics Science Division's Internal Scientist Funding Model. In this study, observations of the Himawari-8 spacecraft, provided by Japanese National Institute of Information and Communications technology was utilized. This dataset was also collected and provided under the Data Integration and Analysis System (DIAS), which was developed and operated by a project supported by the Ministry of Education, Culture, Sports, Science and Technology.

**SUPPLEMENTARY INFORMATION**
File 1: cfg_GCM_EPIC.txt
File 2: cfg_HWO_req.txt